\documentclass[twocolumn,aps,prb,english,twocolumn,epsfig,groupedaddress]{revtex4}
\usepackage{epsfig}
\usepackage{amsmath}
\usepackage[latin1]{inputenc}
\usepackage{latexsym}
\usepackage{ulem}
\begin{document}
\title{Superexchange induced canted ferromagnetism in dilute magnets}
\author{G. Bouzerar,$^{1}$ R. Bouzerar,$^{2}$ and O. C\'epas$^{3}$}
\affiliation{
$1.$ Institut N\'eel, d\'epartement MCBT, 25 avenue des Martyrs, C.N.R.S., B.P. 166 38042 Grenoble Cedex 09, France \\
$2.$ Universit\'e de Picardie Jules Verne (LPMC), 33 rue Saint-Leu, 80039 Amiens Cedex 01,
France. \\
$3.$ Laboratoire de physique th\'eorique de la mati\`ere condens\'ee, C.N.R.S. UMR 7600, Universit\'e
 Pierre-et-Marie-Curie, Paris, France.}

\date{\today}
\begin{abstract}
We argue, in contrast to recent studies, that the antiferromagnetic
  superexchange coupling between nearest neighbour spins does not
  fully destroy the ferromagnetism in dilute magnets with long-ranged
  ferromagnetic couplings. Above a critical coupling, we find a
  \textit{canted} ferromagnetic phase with unsaturated moment. We have
  calculated the transition temperature using a simplified local Random Phase Approximation 
procedure which accounts for the canting. For the dilute magnetic semiconductors,
  such as GaMnAs, using \textit{ab-initio} couplings allows us to predict the existence of a canted
  phase and provide an explanation to the apparent contradictions observed in experimental
  measurements. Finally, we have compared with previous studies that
  used RKKY couplings and reported non-ferromagnetic state when the
  superexchange is too strong. Even in this case the ferromagnetism
  should remain essentially stable in the form of a canted phase.
\end{abstract}
\maketitle

\section{Introduction}

The physics of disordered/dilute magnetic systems has attracted a
considerable interest and attention from both theoreticians and
experimentalists. Among these materials one finds for instance
manganites (LaSrMnO$_3$, LaCaMnO$_3$,..),
\cite{Jin,Schiffer,revue-manga1,revue-manga2} diluted magnetic
semi-conductors as GaMnAs\cite{Ohno} which were widely studied, the 
so-called $d^{0}$ materials (HfO$_{2}$, CaO,...),\cite{d01,d02} the Heusler
alloys as Ni$_{2}$MnSn,\cite{kuebler,Groot} or the double perovkites as
Sr$_2$FeMoO$_6$.\cite{dperov,Sarma} In these materials one of the key
issue is the understanding of the influence of the carrier
(hole/electron) concentration on both magnetic and transport
properties. Indeed, the variation of the carrier concentration often
leads to drastic changes and gives rise to interesting physics. In
particular a competition arises between direct or
superexchange\cite{Anderson} interaction of the localized
magnetic moments and indirect couplings via the itinerant carriers.
In general, the superexchange coupling dominates at low carrier concentration but
is overtaken by the ferromagnetic contribution at higher concentration.
 For example, in manganites (non dilute)
the superexchange coupling competes with the double exchange coupling and leads
to canted ferromagnetic phases.\cite{deGennes,ref1,ref2} However, as
soon as disorder is introduced into the system, new magnetic phases
may appear such as ferromagnetic droplets in a canted
antiferromagnetic matrix as observed in manganites.\cite{ref3,ref4} In
dilute systems, where the probability to have nearest neighbour pairs
is small it is not clear whether the superexchange coupling has the same
effects. In particular, it is not obvious that superexchange alone can eventually
completely destroy the ferromagnetic phase or induce new phases. The
aim of the present study is to focus on this issue.

\begin{figure}[htbp]
\includegraphics[width=6.50cm,angle=-0]{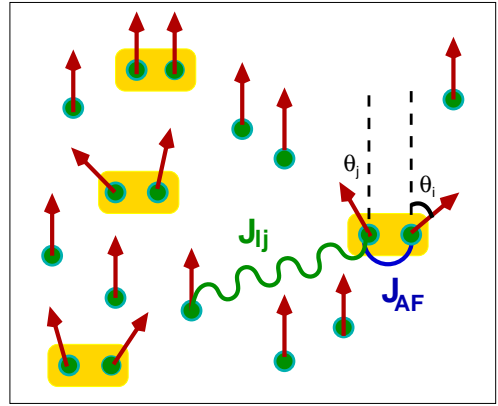}
\vspace{-0.0cm}
\caption{(Color online.) Schematic representation of the canted
ground-state resulting from the competition between the long range ferromagnetic couplings $J_{ij}$ and the
superexchange coupling $J_{AF}$. The spins involved in pairs get canted and the angles $\theta_i$ vary
from spin to spin.}
\label{Fig1}
\end{figure}

In this paper we show that in a dilute system of classical spins the
superexchange competes with the long-ranged ferromagnetic couplings
and favors a canted ferromagnetic phase in part of the phase
diagram (temperature-concentration). However, in contrast to non
dilute materials and double exchange systems, only spins involved in
nearest neighbour pairs get canted (Fig.~\ref{Fig1}). This is
particularly relevant for diluted magnetic semiconductors (DMS). In DMS, the magnetic couplings
are extended and the superexchange dominates at sufficiently low carrier concentration. 
In these materials, a conflict between the measured low temperature total magnetic moment
obtained by SQUID measurements and the density of spins extracted from
X-ray diffraction (XRD) is often observed.\cite{Goennenwein,Korzhavyi,VanEsch,Yu,Potashnik} We shall
see that the existence of a canted phase provides a natural
explanation to the observed disagreement. We also solve a conflict
between recent Monte Carlo simulations\cite{Bergqvist} which found
ferromagnetism in a region where the Self-Consistent-Local RPA
predicted an instability.\cite{Bouzerar1} This instability actually
signals a new phase with unsaturated ferromagnetism, as we shall
see. Because the nature of the ground state was not analyzed, this
conclusion was missed in the Monte Carlo studies which focused on the
amplitude of the Curie temperature only.

The manuscript is organized as follows. In the first part, we will
analyze the effect of the superexchange coupling assuming a simple
model for the extended exchange integrals. In the second part we will
discuss the specific case of GaMnAs where it is known that superexchange coupling
dominates over the indirect ferromagnetic contribution for sufficiently low hole density. 
In this part, for a quantitative study, realistic couplings
will be taken from {\it ab-initio} calculations (TB-LMTO).\cite{Josef}
In the third and last part, we will discuss the case where the
couplings are of the RKKY form, in order to study the competition
between superexchange and frustration effects induced by the oscillating tail. It
will be shown that, in the presence of the superexchange coupling, the stability
region is significantly larger than found in previous
studies.\cite{DasSarma2}

In the following, we will consider the diluted Heisenberg Hamiltonian which reads,
\begin{eqnarray}
H=-\sum_{ij} x_{i} x_{j}J_{ij}\vec{S}_{i}\cdot \vec{S}_{j} + \sum_{\langle ij \rangle}x_{i} x_{j}J_{AF}\vec{S}_{i}\cdot \vec{S}_{j}
\label{Hamiltonian}
\end{eqnarray}
where the random variable $x_{i}$ is 1 if the site is occupied by a
 magnetic impurity (otherwise 0). The total concentration of magnetic
 impurities is $x$. The localized spin $\vec{S}_{i}$ at site $i$ is
 classical ($|\vec{S}_{i}|=1$). The first term corresponds to the
 long-range exchange couplings and the second term is the nearest
 neighbour antiferromagnetic superexchange contribution. Because we will discuss the particular case 
of GaMnAs, for convenience we have performed all the calculations for a fcc lattice.

\section{A simple unfrustrated model}

In this section we consider a simple model where the couplings are
relatively extended but all ferromagnetic $J_{ij}=
J_{0}e^{-\frac{r_{ij}}{\lambda}}$. The parameter $\lambda$ controls
the range of the couplings; but since there is no abrupt cut-off,
there is no strict percolation threshold for the ferromagnetism in this problem. The tail of the couplings always induce a
finite transition temperature. In fact, this model is not so far from
the exchange couplings in III-V diluted magnetic semiconductors such
as Ga$_{1-x}$Mn$_{x}$As or Ga$_{1-x}$Mn$_{x}$N as calculated from
first principles. More realistic couplings will be used in the next
section. Whilst for $J_{AF}=0$ and at $T=0~K$, the ground state is
ferromagnetic and fully saturated (no frustration), we discuss its nature in the
presence of $J_{AF}$. For this purpose, for a given configuration of
disorder (position of the magnetic impurities), we minimize
numerically the total energy associated to the Hamiltonian
(\ref{Hamiltonian}) with respect to the angles
$(\theta_{i},\phi_{i})$. For simplicity, we consider only the case
where the spins are coplanar ($\phi_{i}=0$). The calculations were
performed for systems containing typically 1000 diluted spins. We have
found that beyond a critical value of $J_{AF}$, one pair of nearest
neighbour spins ($\vec{S}_{i}$,$\vec{S}_{j}$) starts to get canted
(see Fig.~\ref{Fig1}) whereas all the other spins remain aligned along
the magnetization axis. As we increase $J_{AF}$ more pairs get canted
but the spins that have no nearest neighbour remains almost parallel
to the z axis. The canting results from the competition between the
local field resulting from the long range couplings and the superexchange contribution of
the nearest neighbour spin. Each spin sees a different environment
and therefore the canting does not occur simultaneously for all the
pairs as we increase $J_{AF}$. For the same reason, the canting angles
are different from spin to spin. For the unpaired spins, however, the
canting angle is very small. This results from two combined
effects. On one hand, for each pair, $\theta_{i}$ is close to $
-\theta_{j}$, so that their resulting transverse field is small. On the
other hand, a given unpaired spin $\vec{S}_k$ experiences the sum of
the transverse fields due to all canted pairs. Because of their random sign, the sum 
averages out to a small value. We can therefore neglect the small canting
angles of the unpaired spins, as shown in Fig.~\ref{Fig1}.

For simple illustration, we recall what happens to a single pair in the
 effective field of the other spins.  The energy
of this pair is $E= J_{AF}\cos(\theta_i-\theta_j)-
 h_i \cos \theta_i-h_j\cos \theta_j $, where $h_i=\sum_{l} J_{il}$ is
 the local field on spin $i$ (for simplicity, we assume $h_{i}=h_{j}=h$ to be the same for both sites, so that $\theta_i=-\theta_j\equiv\theta$).
The minimization gives a canting angle,
\begin{equation}
\cos \theta = \frac{h}{2J_{AF}}
\label{eq0}
\end{equation}
for $J_{AF} \ge h/2$ (and $\theta=0$ otherwise). For $J_{AF}
\rightarrow \infty$, the two spins are anti-aligned and orthogonal to
the other spins: they are effectively decoupled from them. We emphasize that,
in our calculations, we have kept the real local fields $h_i$ that
differ from site to site. In addition we also have a finite
probability to have trimers of spins, quadrimers, etc. so that the
real canting angles are not given by (\ref{eq0}) but are determined
self-consistently.

Now, in order to calculate the critical temperature, we include thermal
fluctuations about the ground state. In the case where the ground
state is fully polarized, it has been shown that the self-consistent
local random phase approximation method
(SC-LRPA)\cite{Bouzerar1,Bouzerar2,RichardPRB} is reliable. When
directly applied to models in presence of superexchange couplings, it has been
argued that ferromagnetism disappears when the nearest neighbour
coupling is dominated by the antiferromagnetic superexchange
contribution.\cite{Bouzerar1,RichardPRB} In fact, the reported
instability turns out to occur at the same critical value at which the
ground-state becomes canted, as found above. The instability
therefore directly reflects the change of ground state and should not
be interpreted as the result of frustration of the long-range
couplings. In order to correct for this, we extend the SC-LRPA to
calculate the critical temperature of a canted state. In principle, as
in the original SC-LRPA one has to start with the equation of motion of the
retarded Green's function, $G_{ij}^{\mu \nu}(\omega)$, where,
\begin{eqnarray}
G_{ij}^{\mu \nu}(\omega)= -i \int^{+\infty}_{-\infty} dt e^{i\omega t} \theta(t) \langle [S_{i}^{\mu}(t),
S_{j}^{\nu}(0)] \rangle
\end{eqnarray}
where $\mu,\nu$ are the spin components.  Because of the canted
ground-state, the decoupling of the equation of motion involves both
the longitudinal $\langle S^{z}_{i}\rangle$ and transverse
magnetizations $\langle S^{x}_{i}\rangle$.  As a consequence, the
transverse Green's function, $G_{ij}^{+-}$, is now coupled to both
$G_{ij}^{--}$ and $G_{ij}^{z-}$.  Since solving these coupled
equations is more involved,\cite{GB-OC} we propose a simplified
\textit{ansatz}: after the
determination of the canting angles $\{\theta_{i}\}$ (for a given
configuration of disorder) we map the canted problem to an effective
fully ferromagnetic one with reduced spin amplitude, $S_{i}
\rightarrow S\cos(\theta_{i})$. Note that, this is equivalent to
replace the couplings by $J_{ij} \rightarrow J_{ij}
\cos(\theta_{i})\cos(\theta_{j})$.  The advantage of this mapping is
that we can use the standard SC-LRPA to calculate the Curie
temperature $T_{C}$, although the ground state is canted.  Note also
that, this mapping is exact in the two limiting cases: (i) small superexchange
coupling and (ii) large $J_{AF}$ limit where the canted pairs become
deconnected from the system. As will be discussed in the following, a
comparison with Monte Carlo results supports this procedure in
the intermediate coupling regime as well.

\begin{figure}[htbp]
\includegraphics[width=7.0cm,angle=-90]{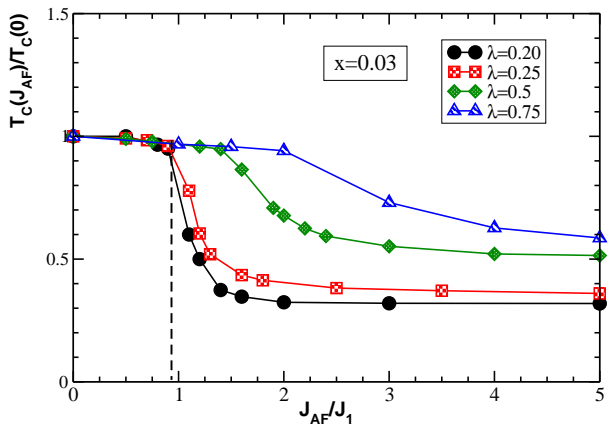}
\vspace{-0.0cm}
\caption{(Color online.) Curie temperature as a function of the
  superexchange coupling strength.  The density of magnetic impurities
  is set to $x=0.03$ and the coupling range $\lambda$ varies from 0.2
  to 0.75. $J_{1}=J_{0}e^{-\frac{a}{\sqrt{2}\lambda}}$ denotes the nearest
  neighbour coupling in the absence of superexchange. The dashed line shows the instability threshold of the simple (non-canted) SC-LRPA calculation.}
\label{Fig2}
\end{figure}

We now come back to the simple model where the long ranged couplings
are defined by $J_{ij}= J_{0}e^{-\frac{r_{ij}}{\lambda}}$. In
Fig.~\ref{Fig2}, we have plotted the Curie temperature as a function
of $J_{AF}$ for a fixed density of magnetic impurities and various
values of the coupling range. First, as long as the superexchange
coupling is smaller than typically the ferromagnetic coupling between
nearest neighbors, ($J_{1} = J_0 e^{- \frac{a}{\sqrt{2} \lambda}}$),
$T_{C}$ is almost insensitive to $J_{AF}$. This is in agreement with
previous observation that in the diluted regime, the Curie temperature
is controlled by couplings corresponding roughly to the average
distance between the magnetic impurities, $x^{-1/3}$. When $J_{AF}$ is
increased, there is a critical value above which the ground state gets
canted (the critical value increases with $\lambda$). When this
happens, we observe a reduction of $T_{C}$, and then a saturation to a
finite value for strong $J_{AF}$. Let us discuss the limit of strong
$J_{AF}$.  The saturation of $T_C$ corresponds to the regime where
nearest neighbour spins are orthogonal to the other spins. The
saturated value can be viewed as that of a system of $x_{eff}$ spins
(interacting with ferromagnetic interactions) in which all pairs have
been removed. For example, on the fcc lattice, for $x=0.05$ the
concentration of spins involved in pairs is $0.0225$, or
$x_{eff}=0.55~x$. The new characteristic distance between remaining
impurities is $x_{eff}^{-1/3}$ and has to be compared to the coupling
range $\lambda$, which controls whether the impurities ``percolate''.
When reducing $\lambda$, the ratio of the saturating value over
$T_{C}(0)$ gets quite large (see Fig.~\ref{Fig2}) because we approach
the regime of non-``percolation'', where the remaining impurities get
more weakly coupled. To conclude this paragraph, we observe a smooth
crossover in the Curie temperature between a weak-coupling regime
where $J_{AF}$ has no effect and a strong-coupling regime where it is
equivalent to remove all spins which are coupled by $J_{AF}$. This is
true for a concentration of impurities small enough and we now
investigate the effect of varying the concentration.
\begin{figure}[htbp]
\includegraphics[width=7.0cm,angle=-90]{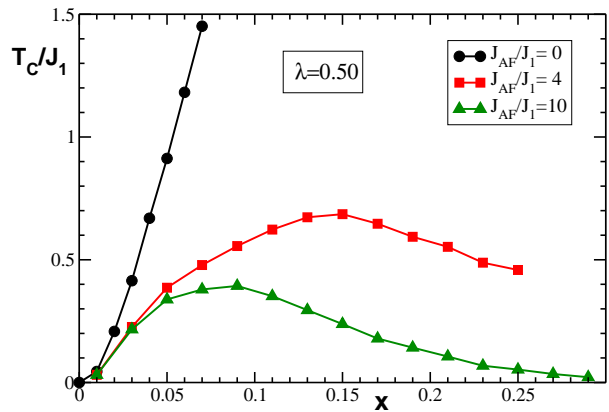}
\vspace{-0.0cm}
\caption{(Color online.) 
Curie temperature (in units of $J_{1}$) as a function of the magnetic impurity concentration $x$ for different values of the superexchange strength. The range of the couplings, $\lambda$, is fixed to 0.50. }
\label{Fig3}
\end{figure}

In Fig.~\ref{Fig3}, we have plotted the Curie temperature as a
function of the impurity concentration, $x$, for a fixed value of the
parameter $\lambda$ and different values of $J_{AF}$. In the absence
of superexchange coupling, we observe a strong increase of $T_{C}$
with $x$. For $J_{AF}=4J_{1}$, the Curie temperature is strongly
reduced and exhibits a maximum at $x \approx 0.15$. The maximum
reduces to $x \approx 0.08$ and is more pronounced for
$J_{AF}=10J_{1}$. Above this value the Curie temperature decreases
strongly to eventually vanish at $x\approx 0.30$. The presence of a
maximum can be understood as resulting from the competition between
two effects. As we increase the density of magnetic impurities, $x$,
the local fields increase (the impurities interact more strongly
because they get closer). At the same time the probability of nearest
neighbor spins increases also.  Since nearest neighbor spins get
canted the local fields they create on the other spins is
reduced (and eventually vanish in the limit of infinite $J_{AF}$), so
that the number of magnetically-active spins (as far as ferromagnetism
is concerned) is effectively reduced. Note that for strong $J_{AF}$,
one expects a site percolation which for nearest neighbour coupling on fcc lattice occurs at $x_{C}\approx 0.20$.
Beyond this critical value, the phase should be of N\'eel type. One would need
  to include all fluctuations in order to calculate its critical N\'eel
  temperature, a problem beyond the scope of the present paper, since we focus on the 
ferromagnetic phase only.

\section{Realistic couplings: the case of $\mbox{GaMnAs}$}

Let us now discuss the case of the widely studied diluted III-V
magnetic semiconductor Ga$_{1-x}$Mn$_{x}$As. First, we remind that the
substitution of Ga$^{3+}$ by Mn$^{2+}$ introduces a localized spin
S=5/2 and a hole in the valence-band (more precisely in the impurity
band).

During the molecular beam epitaxy (MBE) growth of the samples, there
are additional defects which appear, namely As anti-sites
($\mbox{As}_{\mbox{Ga}}$) which substitute the Ga sites.  Arsenic
antisites formation is one of the main mechanism for compensation in
diluted magnetic semiconductors.  They lead to the reduction of the
density of carriers which in turn reduces the strength of the magnetic
couplings and eventually the Curie temperature. In term of holes,
$\mbox{As}_{\mbox{Ga}}$ is a double acceptor (double donor of
electrons). If $y$ denotes the density of As antisites and $x$ the
density of Mn$^{2+}$ then the density of holes is $n_{h}=x-2y$.  The
reduction of the carrier density via $\mbox{As}_{\mbox{Ga}}$ not only
affects the long range couplings but also allows to "tune" the superexchange
coupling \cite{Si,Larson}, and thus provides a way to test the ideas
developped above. Indeed, beyond a certain concentration of
As$_{\mbox{Ga}}$ the superexchange mechanism dominates the nearest
neighbour coupling (see Figure 5 in [\onlinecite{Bergqvist}]). This is
clear from {\it ab initio} studies, where the couplings have been
calculated without any adjustable parameters (Tight-Binding Linear
Muffin Tin Orbitals), and subsequently used in several
publications.\cite{Bouzerar1,Bouzerar2,Bergqvist}
\begin{figure}[tbp]
\vspace{-0.0cm}
\includegraphics[width=7.0cm,angle=-90]{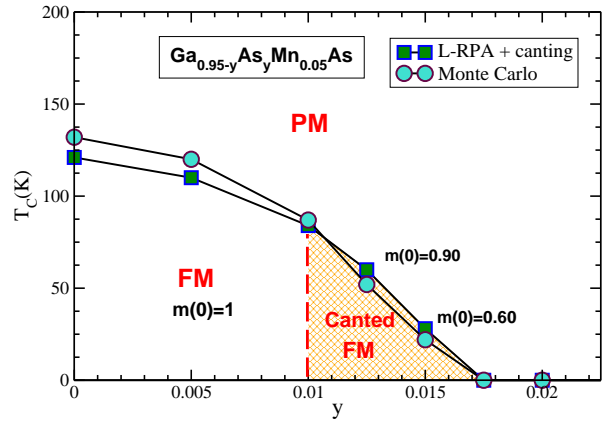}
\caption{(Color online.) Calculated phase diagram (temperature-antisite concentration) for Ga$_{1-x-y}$As$_{y}$Mn$_{x}$As indicating the predicted \textit{canted} phase. The antisite concentration $y$ allows to tune the carrier concentration. The density of magnetic impurity $x$ is fixed to $0.05$.  The transition temperature is calculated with modified LRPA, and compared with Monte Carlo (from ref. [\onlinecite{Bergqvist}]), using the same set of \textit{ab-initio} couplings. $m(0)$ is the total magnetization per spin at zero temperature.
 }
\label{Fig4}
\end{figure}

In Fig.~\ref{Fig4}, using SC-LRPA in the same way as described in section II, and using
ab-initio couplings, we calculate the predicted phase diagram for
Ga$_{1-x-y}$As$_{y}$Mn$_{x}$As (temperature - density of As
antisites). By energy minimization, we have found a wide region of the phase diagram ($0.01 < y < 0.0175$) where the
ground state is not fully ferromagnetic but canted. This is the
consequence of the superexchange coupling $J_{AF}$ that increases when
the antisite concentration gets larger.  As far as the transition
temperature is concerned, we stress that the values obtained are in
very good agreement with that of Monte Carlo
simulations.\cite{Bergqvist} Although the nature of the phase was not discussed
in their study, this validates the simplified treatment of
the canting of the ground-state. Note that, both Monte Carlo
simulations and SC-LRPA were performed (i) with the same exchange
couplings and (ii) with the same number of shells (approximately
20). Note also that the values of the Curie temperature in the fully
polarized ferromagnetic phase ($y \le 0.01$) are a little
smaller here than those published in ref. [\onlinecite{Bouzerar2}] because
the number of shells is smaller. More importantly, in this study, the
Curie temperature vanished abruptly beyond $y=0.01$ (see
Fig.~1 of ref. [\onlinecite{Bouzerar2}]). This instability in fact
signals the occurence of a new phase, that we identify as a
canted phase. Thus, in contrast to what was claimed before, the ferromagnetism
survives down to much smaller concentrations of carriers than anticipated (up to $y\approx
0.0175$), but is non-saturated. Beyond this concentration, the
canted phase disappears because the long-range couplings also become
antiferromagnetic and thus introduce real frustration into the
system. Because of that, for $y\ge 0.0175$ the ground-state is expected
to be a spin-glass.
 
To characterize the ground state, we have also calculated the total
magnetization per spin $m(0)$ at $T=0~K$. For $y \le 0.01$ the
magnetization is $m(0)=1$ (by definition). As we enter the canted
ferromagnetic phase the total magnetization starts to reduce
significantly. For $y=0.0125$ the magnetization is already reduced by
10\% and for $y=0.015$ it is only $m(0)\approx 0.60$ which is very close to the lower bound obtained by removing all pairs, $m(0) \ge x_{eff}/x =0.55$ for $x= 5\%$.
\begin{figure}[tbp]\hspace{-0.2cm}
\includegraphics[width=7.0cm,angle=-90]{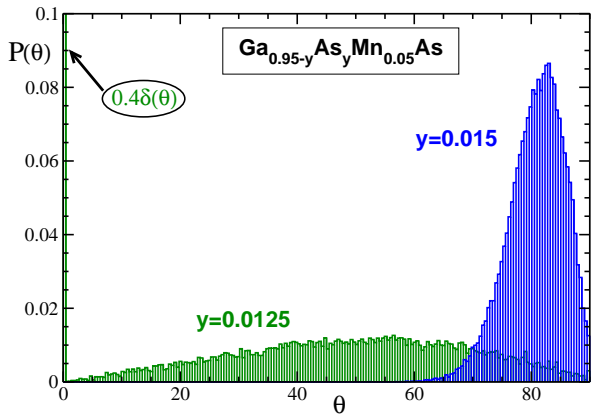}
\vspace{-0.0cm}
\caption{(Color online.) 
Distribution P($\theta$) of the canting angle of the nearest neighbour pairs in 
Ga$_{1-x-y}$As$_{y}$Mn$_{x}$As for two different values of the concentration of As anti-sites y.
The density of Mn$^{2+}$ is $x=0.05$.
}
\label{Fig5}
\end{figure}
In order to get an idea of how the magnetization changes from site to site,
we have plotted the distribution of canting angles in Fig.~\ref{Fig5}. Note that the distribution is given 
for spins having at least one nearest neighbor. We observe for $y=0.0125$ that
approximately 40\% of the spins are still not canted (delta peak at
$\theta=0$).The distribution of angles is very broad with a maximum at
about $50^o$. On the other hand, for $y= 0.015$ we observe a
strong change in the distribution. In the latter case, all spins are
canted, and the distribution peaked at about $\theta \approx 80^{o}$
is more narrow than that of $y=0.0125$.

Let us now discuss the relation between our calculations and
experimental data. It is often seen in the literature that the
measurement of the bulk magnetization (by SQUID) is different from the
magnetization expected from the determination of the Mn density from
XRD measurement, assuming a fully polarized
ferromagnet.\cite{Goennenwein,Korzhavyi,VanEsch,Yu,Potashnik} The
direct measurement often leads to much smaller values. Furthermore, it
is also seen that the magnetization strongly changes after annealing
of as grown samples. During annealing, the magnetic impurities are
redistributed in the sample, which becomes more
homogeneous.\cite{Kirby} In a recent study,\cite{Bouzerar2} it was
shown that one could explain the effect of different annealing
treatments\cite{Edmonds} by the existence and the rearrangement of
interstitial Mn defects (Mn$_{I}$).  Indeed Mn$_{I}$ is a defect that
preferentially sits near a Mn ion (which substitutes Ga) and is
coupled antiferromagnetically to it.\cite{Blinowski,Masek} This leads
to the formation of a local singlet state for the dimer of Mn and
therefore reduces the number of magnetically active Mn, and hence the
total magnetization.  However, it is now possible to control the
density of carriers by chemical hydrogenation of the
samples.\cite{Goennenwein,Thevenard} In this process, it is believed
that the density of Mn and the density of defects Mn$_I$ do not
change. Therefore, if interstitial Mn$_I$ defects were the main source
of reducing the bulk magnetization $m(0)$, one would expect $m(0)$ to
remain the same for all these hydrogenated samples. This is in
contradication with the measurements that indicate that the samples
with lowest carrier density (insulating or very dirty metallic
behavior) have a much smaller $m(0)$. For instance, in Fig.~3 of ref.
[\onlinecite{Thevenard}] at small fields ($H=500~\mbox{Oe}$), we
observe that the magnetization is about \textit{two} times smaller for
hydrogenated sample with the lowest $T_C$, compared with the reference
sample with no hydrogenation. This is very hard to reconcile with the
presence of Mn$_I$ because this would require a large number of such
defects. Our study points out a different compensation mechanism, that
must be at play once the coupling between nearest neighbours is
antiferromagnetic (as evidenced from \textit{ab initio} studies). As
we said before, the number of pairs of nearest neighbors is 
large at $x=5\%$, so the reduction of $m(0)$ is already large
without having to invoke a large number of Mn$_I$. We therefore argue
that the reduction of the total magnetization is due to the canting of
the pairs which occurs when the density of carriers becomes small
enough, and we suggest to reanalyze the experimental data on the basis
of the present work.

\section{Model with RKKY oscillations}

In the last paragraph we discuss another interesting case, the
interplay between frustration resulting from the RKKY couplings and
the superexchange. In order to compare with previous studies, we
assume the long range couplings to be given by $J_{ij}=
J_{0}e^{-\frac{r}{\lambda}}F(k_{F}r)$, where
$F(k_{F}r)=(k_{F}r)\cos(k_{F}r)/(r/r_{0})^{3}$ and $r=|r_{i}-r_{j}|$,
$r_{0}$ is the nearest neighbour distance . The parameter is an
effective Fermi vector $k_{F}$ which is determined by the density of
carriers. Note that this study is also motivated by the fact that RKKY
couplings are often used to study the ferromagnetism in diluted
magnetic semiconductors,\cite{DasSarma1,Dietl} although it was shown
that they are inappropriate.\cite{RichardPRB} In previous studies, it was argued, in particular, that the
stability region for ferromagnetism was very narrow, upon increasing $J_{AF}$,\cite{RichardPRB}
a point that was  reaffirmed later using Monte Carlo
simulations.\cite{DasSarma2} We now argue that the stability region is in fact wider thanks to unsaturated phases.

\begin{figure}[tbp]\hspace{-0.2cm}
\includegraphics[width=7.0cm,angle=-90]{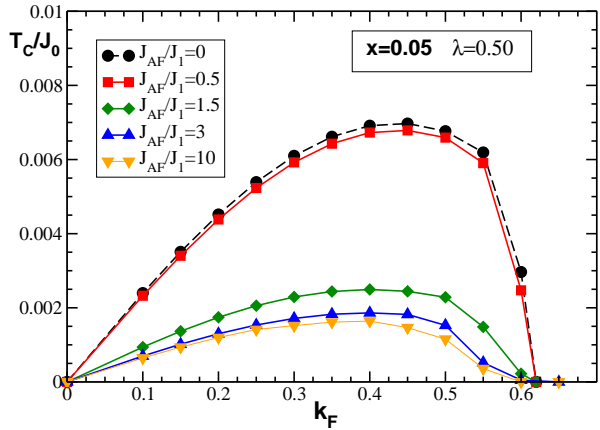}
\vspace{-0.0cm}
\caption{(Color online.) 
Curie temperature for the ``RKKY'' model as a function of $k_{F}$ for different values of the superexchange coupling $J_{AF}$. The density of impurity is set to $x=0.05$ and the parameter $\lambda=0.50$. The calculations are performed on the fcc lattice.
}
\label{Fig6}
\end{figure}

In Fig.~\ref{Fig6} we have plotted the Curie temperature as a function
of $k_{F}$ for the RKKY-like model defined above. In the absence of superexchange
coupling we observe that the Curie temperature (solid circles) exhibits a maximum and
vanishes above $k_{F} \approx 0.60$ because of the oscillations in the couplings. This
is in agreement with ref. [\onlinecite{RichardPRB}]. However, when we
switch on the superexchange coupling, we observe that the Curie temperature is
reduced but does not vanish. This is in contrast to what was published
previously where the ferromagnetism was apparently suppressed by the
superexchange coupling. The reason for this discrepancy is as discussed
previously, the occurrence of a new phase. Indeed previous
calculations were done without including the canting. Again, the
disappearance of $T_{C}$ (Fig.~5 of ref. [\onlinecite{RichardPRB}])
reflects only the change in the nature of the ground state.  As we
increase $J_{AF}$ further, we observe a saturation in the Curie
temperature and almost an unchanged region of stability of the
ferromagnetic phase.  These results are also in contradiction with
those of ref. [\onlinecite{DasSarma2}]. Indeed, it is shown in this
paper that the ferromagnetic phase is similarly suppressed (see Fig.~3
of ref. [\onlinecite{DasSarma2}]). The drastic reduction of the
ferromagnetic phase occurs also at low carrier concentration, although
the frustration effects due to the long range couplings are very small
in this region. Let us discuss the origin of the discrepancy.  In the
absence of $J_{AF}$, Monte Carlo\cite{DasSarma2} and
SC-LRPA\cite{RichardPRB} give the same results (see Fig.~1 and 4 of
ref. [\onlinecite{RichardPRB}] and Fig.~1c of
ref. [\onlinecite{DasSarma2}]). The effects of frustration are
therefore properly handled by SC-LRPA, and thus it cannot be the
source of the disagreement.  In addition, the comparison between the
Monte Carlo simulations of [\onlinecite{Bergqvist}] with the modified
SC-LRPA clearly shows that the effects of the superexchange are also properly
treated. We suggest that the discrepancy comes from the existence
of the canted phase that was missed before. It would be of interest to
clarify this issue.

\section{Conclusion}

In conclusion, we have shown that the competition between
nearest-neighbour antiferromagnetic superexchange coupling and
long-range ferromagnetic couplings gives rise to a canted
ferromagnetic phase in dilute magnets.  We emphasize that short-range
and long-range competing interactions play quite different roles. In
the later case, the oscillating tail of the RKKY interaction, for
instance, introduces frustration at long distance and consequently
reduces the stability region of the ferromagnetic phase. In the first
case, however, when superexchange is added, pairs of spins get canted
but the stability region remains weakly affected, even at strong
coupling. The Curie temperature of the canted phase is reduced simply
because the canting weakens the internal fields. 
More generally, in random dilute systems, competing couplings of a range
much shorter than the typical inter-impurity distance should lead only
to local spin reorientations, and will not affect the long-range
properties. Applying these ideas to GaMnAs, we have predicted the existence of a
canted phase in the phase diagram. We have calculated its critical
temperature using a modified local RPA approach that was found to be
reliable by comparison with Monte Carlo simulations using the same
{\it ab initio} couplings. The existence of this phase provides a
simple explanation to recent experiments in diluted magnetic
semiconductors, where the bulk magnetization was found to be smaller
than the saturation value; without having to invoke a large number of
compensating defects in the samples. It would be of great interest to
check by local probes whether the ground state is indeed canted in the
range of concentrations we have predicted.

\acknowledgments

We would like thank E. Kats for the hospitality at ILL.

\end{document}